# Monitoring field soil suction using a miniature tensiometer

*Yu-Jun Cui,[1] Anh-Minh Tang,[2] Altin Theodore Mantho,[2] Emmanuel De Laure[2]*

**ABSTRACT:** An experimental device was developed to monitor the field soil suction using miniature tensiometer. This device consists of a double tube system that ensures a good contact between the tensiometer and the soil surface at the bottom of the testing borehole. This system also ensures the tensiometer periodical retrieving without disturbing the surrounding soil. This device was used to monitor the soil suction at the site of Boissy-le-Châtel, France. The measurement was performed at two depths (25 and 45 cm) during two months (May and June 2004). The recorded suction data are analyzed by comparing with the volumetric water content data recorded using TDR (Time Domain Reflectometer) probes as well as the meteorological data. A good agreement between these results was observed, showing a satisfactory performance of the developed device.

**KEYWORDS:** Field monitoring, suction, miniature tensiometer, TDR, volumetric water content, water retention curve.

---

[1] Professor, Ecole Nationale des Ponts et Chaussees, (CERMES), 6 et 8 av. Blaise Pascal, 77455 MARNE-LA-VALLEE CEDEX 2, France. Email: cui@cermes.enpc.fr.

[2] Ecole Nationale des Ponts et Chaussees.

**Introduction**

Unsaturated soil mechanics is often applied to geotechnical problems such as embankments, dams, pavements, foundations, landfills, slopes, nuclear waste disposals, etc. Analyzing such problems requires information about soil suction variations. This explains why significant effort has been made from all over the world on suction measurement techniques under field conditions. Rahardjo and Leong (2006) presented several devices and techniques that are used for *in situ* soil suction measurement. Jet fill tensiometer and small-tip tensiometer are often used to instrument soil slopes (Rahardjo et al. 2005). Konrad and Ayad (1997) used small-tip tensiometer to observe the desiccation of sensitive clay in field. In general, the soil suction range of these tensiometers is below 100 kPa. Ng et al. (2003) used thermal conductivity sensor to measure higher suctions in an unsaturated swelling soil slope. The highest suction recorded reached 900 kPa. Although the suction range was enlarged using thermal conductivity sensor, hysteresis phenomenon and long time for stabilization are still the drawbacks of this technique (Nichol et al. 2003). Wray et al. (2005) reported field data from three test sites where soil suction was measured using thermocouple psychrometer. The suction range for the involved soils varied from 100 kPa to 10 MPa. According to Andraski and Scanlon (2002), the monitoring of soil suction with psychrometer technique in the near surface zone (less than 0.4 m depth) could be affected by the temperature variations. However, as Cui et al. (2005) reported, in practice only the suction in the near surface zone varies significantly under soil-atmosphere interaction.

Miniature tensiometer has been widely employed in laboratory tests to measure the soil matric suction (Tarantino and Tombolato 2005, Singh and Kuriyan 2003). It has been observed that the initial good saturation of tensiometer is essential to obtain a maximum measured suction;

otherwise, the cavitation phenomenon can occur before the real soil suction is captured. Two techniques are often used to ensure a satisfactory saturation (Take and Bolton 2003): i) initial water percolation under vacuum, ii) cyclic pre-pressurization (high pressure application and cavitation under high suction).

In the present work, a miniature tensiometer was used to monitor the field suction in the near surface zone (at 25 and 45 cm depth). An experimental procedure was developed to install the miniature tensiometer in field. The variation of soil suction during 2 months was interpreted using the recorded meteorological data and the water volumetric content changes measured using TDR probes.

**Field conditions and experimental devices**

The field monitoring test was performed at a location near the village of Boissy-le-Châtel, France, which is located at about 50 km east of Paris, in the South of the Orgeval basin, at an altitude of 133 m above sea level. Since 1996, meteorological data and soil volumetric water content changes have been recorded by CEMAGREF, the French public organization devoted to agricultural and environmental engineering research. More details on this instrumented field site can be found in Cui et al. (2005). The soil can be classified as a silty soil and its Atterberg limits are: plastic limit $PL = 20$ %, liquid limit $LL = 29$ %.

The meteorological data and soil volumetric water content recorded during May and June 2004 are presented in Figure 1. Four main rainfall periods can be observed: (1) from May 4 to May 8; (2) from May 30 to June 1; (3) on June 11; (4) from June 19 to June 22. The daily mean wind speed varies between 2 and 18 km/h (data from May 18[th] to June 10[th] were not recorded).

The global solar radiation ($R$) decreases during the rainfall periods (down to 90 W/m$^2$ on May 9) and increases after each rainfall period (up to 1300 W/m$^2$) on June 6. On the other hand, the air relative humidity ($RH$) increases during the rainfall periods (up to 100 %) and varies between 60 and 80 % in other days. As May and June correspond to the beginning of the summer, it can be observed that the air temperature ($T$) trends to rise and soil water content trends to decrease.

The schematic diagram of the miniature tensiometer is presented in Figure 2$a$. The high-air entry value (HAEV) ceramic stone (1.5 MPa) is stuck on the tensiometer body in stainless steel, with epoxy glue. The water reservoir volume is minimized with a distance of 0.1 mm between the porous stone and the diaphragm. The strain gauges that are stuck on the other side of the diaphragm allowed the monitoring of water pressure applied on the diaphragm after a preliminary calibration.

Prior to use, the tensiometers were saturated by pre-pressurizing with de-aired water at high pressure (until 4 MPa). They were then calibrated in the positive pressures rang and the calibration curve in the negative pressures was extrapolated. All the calibration tests were performed at temperature controlled laboratory conditions, at 20 °C. Similar procedure was described by Tarantino and Mongiovi (2001). In order to estimate the effect of temperature on the response of tensiometer, the tensiometers were calibrated at constant pressure and changing temperature (from 20 °C to 32 °C). It was observed that the pressure given by the tensiometer increases with the temperature rise, at a rate of 1 kPa/°C. This value is close to that obtained by Hoffmann et al. (2006); it was used to make corrections of the recorded suction at Boissy-le-Châtel site.

The schematic diagram of the in-situ suction measurement system is showed in Figure 2$b$. This system consists of two PVC (Polyvinyl chloride) tubes (see also Figure 2$c$). The outer tube (63 mm in exterior diameter) is equipped with a metallic leading edge at the lower end. This facilitates insertion of the tube in soil and the centering of the inner tube. Several O-rings are installed to make the system waterproof.

To install this system in field, a hole is first bored by pushing the outer PVC tube, which is equipped with metallic leading edge, in the soil. When the tube reaches the depth planned for suction monitoring, the soil surface in the bottom of the hole is leveled and cleaned. To provide better contact between the soil and the outer tube, the tube is coated within a thin layer of soil paste. This layer would prevent any "short-circuit" infiltration from the ground surface along the tube during rainfall. Finally, the $\phi$63 mm support is installed to fix the outer tube.

The procedure of tensiometer installation is described as follows:

- fix the tensiometer, which is previously saturated in laboratory, to the inner tube;

- coat a thin layer of the soil paste on the surface of the tensiometer to prevent any desaturation of the porous stone during installation;

- install the inner tube until it reaches the desired depth where the tensiometer is in contact with the soil surface;

- install the $\phi$50 mm support and the shutter support.

The TDR probe contains three waveguide stainless steel rods; each being 20 cm long and 0.3 cm in diameter.

**Experimental results**

The field suction was monitored at two depths (25 and 45 cm) during May and June 2004 and results are presented in Figure 3*a*. It can be observed that the suction values increase during this period. This is consistent with air temperature and soil water content changes (Figure 1).

To prevent the cavitation due to air diffusion from soil to water reservoir through the porous stone, the tensiometer was changed every two or three weeks for re-saturation in laboratory. The changing process lasted about half an hour and resulted in a suction reduction. In Figure 3*a*, the installations of tensiometer correspond to the points where the suction drops down to zero.

During these two months, four rainfall periods were recorded (Figure 3*b*). Except the last rainfall period from June 19 to June 22, the rainfall periods induced a slight reduction of suction at the two depths. The results are zoomed in Figure 3*c,d,e* where the variation of suction is presented with respect to time. During the first period (from May 4 to May 8, Figure 3*c*), a suction reduction of 8 kPa was recorded at the two depths. In addition, this reduction took place at almost the same time. On the contrary, during the other rainfall periods (from May 30 to June 1, Figure 3*d*, and on June 11, Figure 3*e*), the suction reduction at 25 cm was less significant than at 45 cm: the rainfall on May 30 caused a suction reduction of 15 kPa at 45 cm depth and 5 kPa at 25 cm depth; the rainfall on June 1 induced a suction reduction of 5 kPa at 45 cm depth but an insignificant change at 25 cm depth; the rainfall on June 11 resulted in a suction decrease of 25 kPa at 45 cm depth and 8 kPa only at 25 cm depth.

**Discussion**

It can be observed that in low suction range (20 – 160 kPa), the tensiometer can be used to monitor the suction changes during more than two weeks without cavitation. After the last changes of tensiometers on June 7 (45 cm) and June 10 (25 cm), the tensiometers were left in the field for three weeks. Despite this longer duration, no cavitation was observed. Nevertheless, the authors expected that with higher field suction, this duration would be shorter because of a probable more significant air diffusion from soil to the water reservoir of tensiometer. Indeed, the periodical changes of tensiometers should be more frequent. Tarantino and Mongiovi (2001) observed that cavitation occurred after 4 days at 1275 kPa suction when using this type of tensiometer.

A daily fluctuation of the suction values at the two depths can be observed in Figure 3*d*. The fluctuation at deeper level (45 cm) seems to be smaller than at near surface level (25 cm). That can be attributed to the daily fluctuation of air temperature. Indeed, the soil at shallower level is more influenced by climate than at deeper level. For example, during May 31, the suction at 25 cm depth varied from 70 kPa to 75 kPa when air temperature varied from 7 °C to 20 °C. Assuming that the soil temperature at 25 cm depth is equal to the air temperature, the corresponding value of $\Delta \log s / \Delta T$ is $-2.3 \times 10^{-3}$ (1/°C). This value is in the same order of magnitude with the value estimated by Tang and Cui (2005) by considering the interfacial tension changes with temperature, $\Delta \log s / \Delta T = -1.1 \times 10^{-3}$ (1/°C).

For further discussion, the results measured by tensiometer and by TDR probes are plotted in Figure 4. These are well known water retention curves. It appears that no unique relationship exists for the two depths. In addition, for each depth, the fluctuation in suction is quite significant, which can be related to the changes in climatic conditions, especially in terms of rainfalls. As pointed out by several researchers (Pham et al. 2005) the water retention curve depends on soil suction history: because of the well known hysteresis phenomenon there are two bounding curves which delimit a zone where any changes in water content with suction changes are possible. The upper bounding curve corresponds to the drying curve from fully saturated state, whereas the lower one corresponds to the wetting curve from a dry state at residual water content. In Figure 4, the two bounding curves are plotted.

The hysteresis phenomenon can be used to explain the suction changes observed during the rainfall periods (Figure 3*a,b*): it was observed that for the two rainfalls on May 30 and June 11 the deeper tensiometer (45 cm depth) showed a more significant suction decrease than the shallow tensiometer (25 cm depth). From Figure 4, it can be seen that the water retention curve at 45 cm depth situates above the curve at 25 cm depth. Thus, the curve at 45 cm depth is closer to the bounding drying curve and the curve at 25 cm depth is closer to the bounding wetting curve. During rainfall, as both of the two curves tend to rejoin the bounding wetting path, the resulted suction decrease at 45 cm depth must be more significant than that at 25 cm depth.

**Conclusions**

A miniature tensiometer was used to monitor *in situ* soil suction at the site of Boissy-le-Châtel at two depths (25 and 45 cm). An experimental device was developed that allowed a periodical changing of the tensiometer for re-saturation without disturbing the soil. A changing frequency of two weeks was generally applied to avoid tensiometer cavitation. It was observed that even with a longer monitoring duration of three weeks, the tensiometers did not cavitate with a suction ranging from 20 to 160 kPa. However, it is believed that in case of higher field suctions, the monitoring duration would be shorter because of a probable more significant air diffusion from soil to the water reservoir of tensiometer.

Data recorded during two months (May and June 2004) are found to be in good agreement with meteorological conditions and *in situ* water content measurement by TDR probes. The suction tended to increase, in concordance with the air temperature rise and the soil water content decrease. In the near surface zone, the suction seems to be influenced by temperature changes, at a rate: $\Delta \log s / \Delta T = -2.3 \times 10^{-3}$ (1/°C).

Abrupt suction decreases were observed during rainfall periods. In addition, the decrease was different for the two considered depth; the decrease was more significant at 45 cm depth.

Hysteresis seems to be the origin of this difference. Indeed, because during the monitoring period the retention curve at 25 cm depth is closer to the bounding wetting curve than that at 45 cm depth, any rainfall would lead to a more significant suction decrease at 45 cm depth.

**Acknowledgement**

The authors are grateful to CEMAGREF for providing data about climate condition and soil volumetric water content, as well as for making easier the suction monitoring, at Boissy-le-Châtel site.

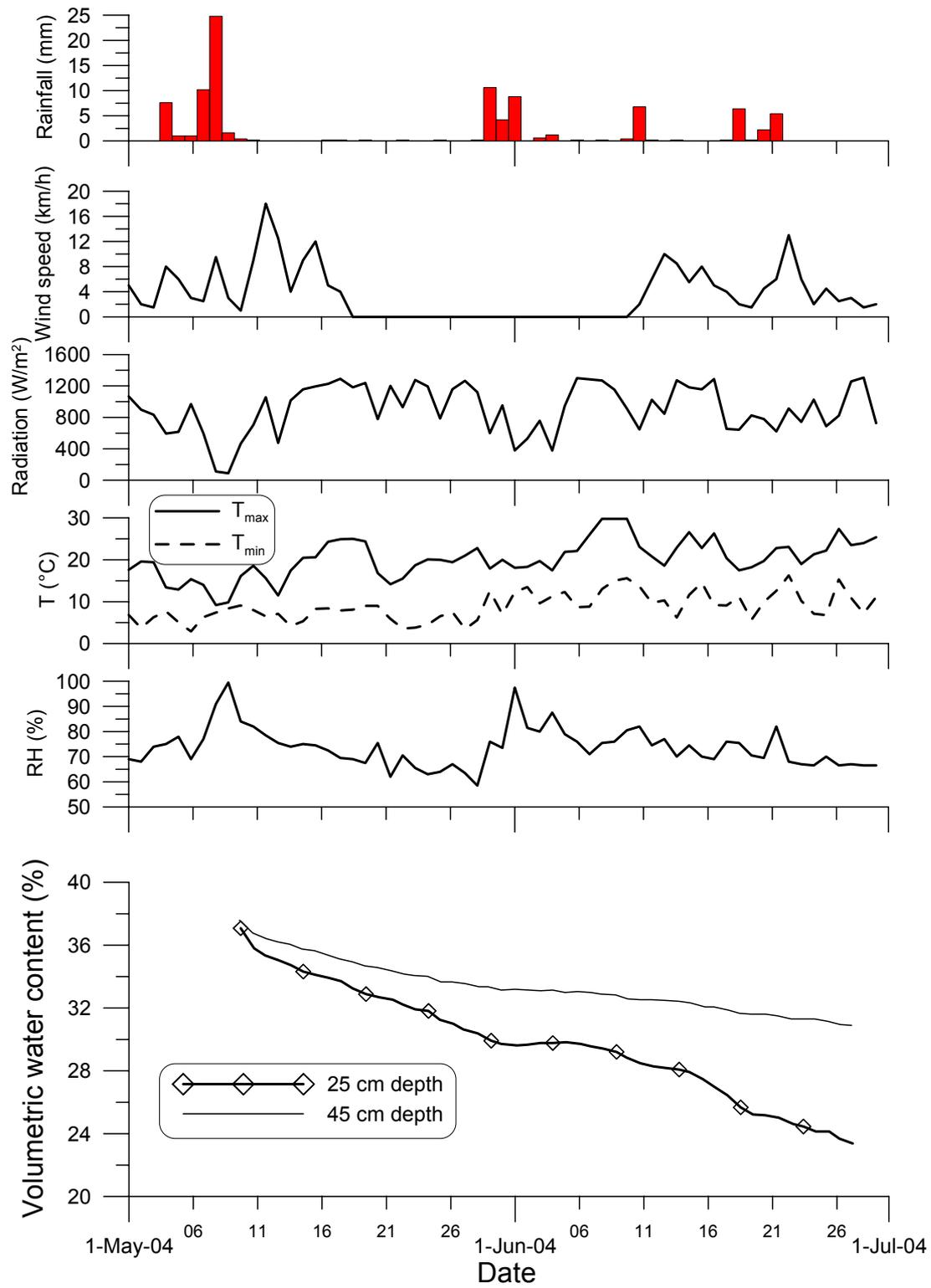

Figure 1. Meteorological data and soil water content during May and June 2004.

(a)

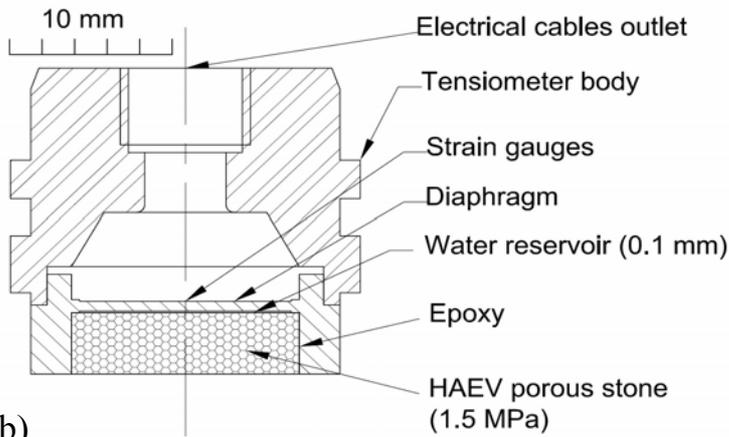

(b)

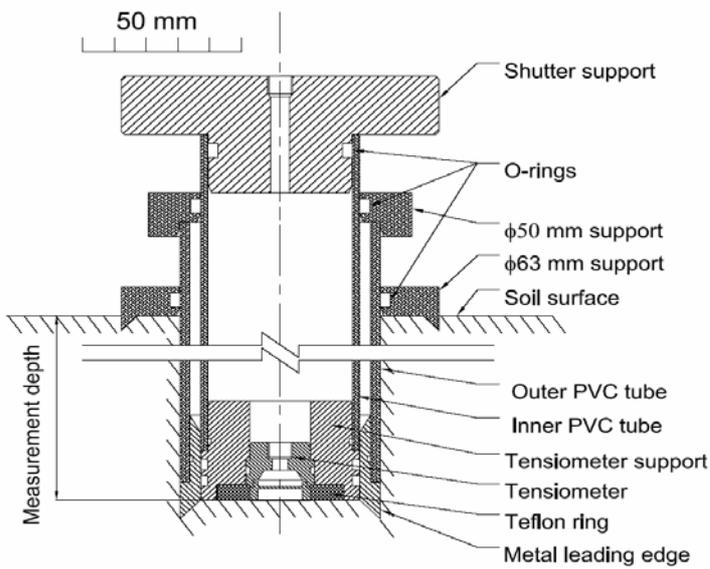

(c)

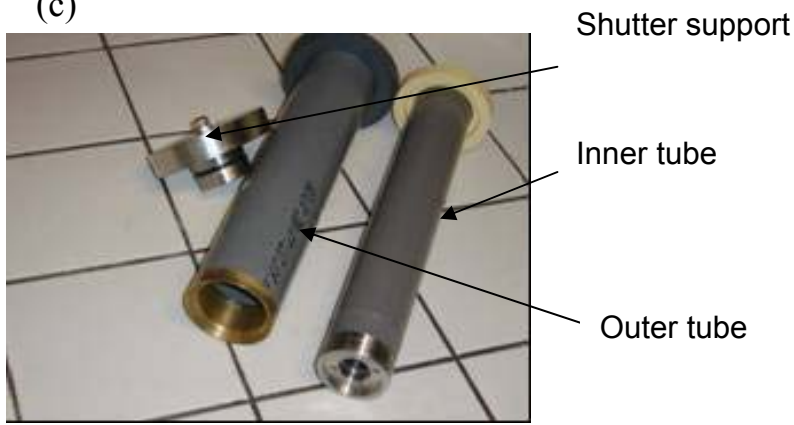

**Figure 2. Schematic layout of miniature tensiometer (a), the in-situ suction measurement system (b), and picture of the inner tube and the outer tube (c).**

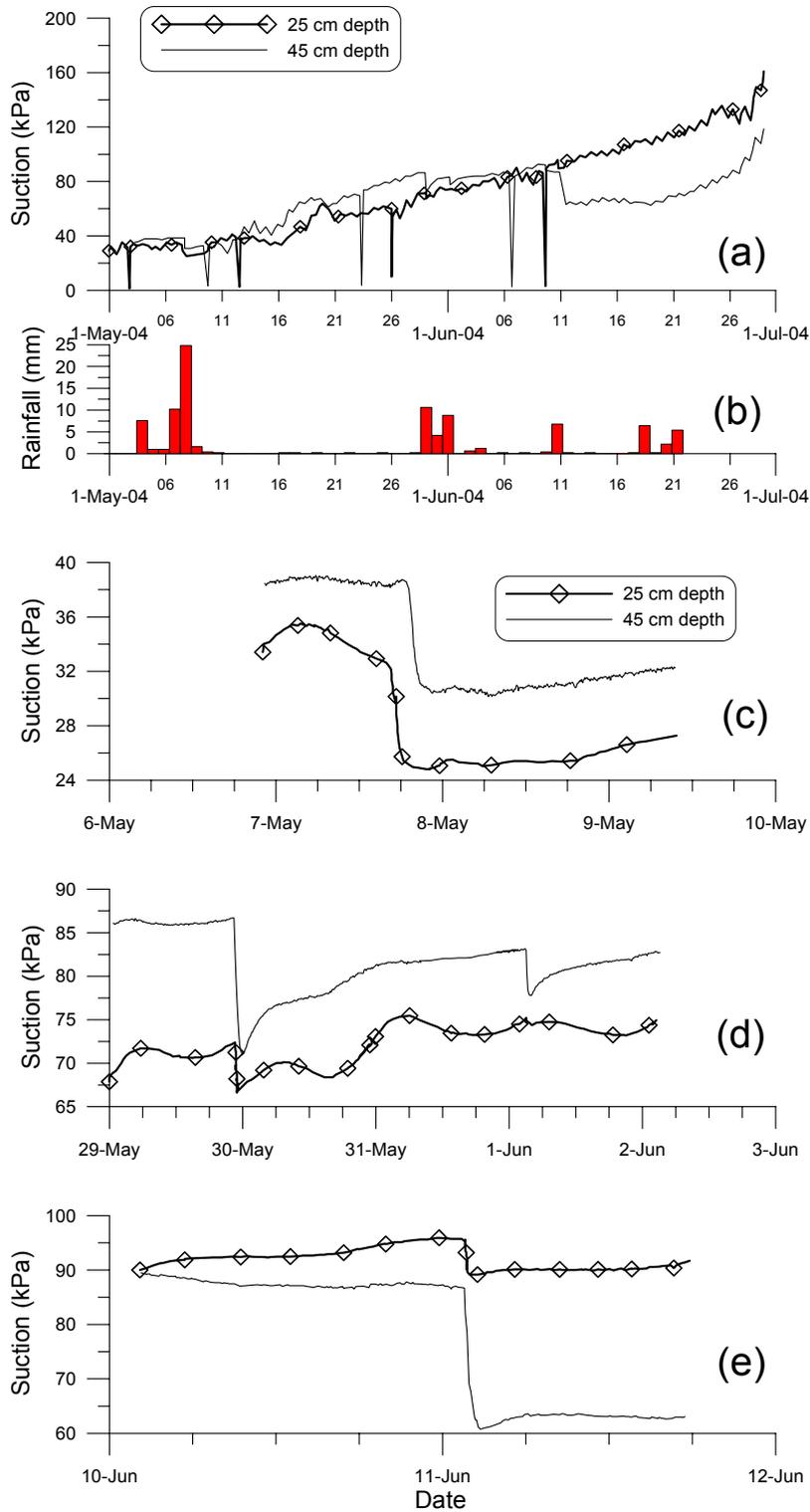

**Figure 3. Soil suction versus time (a) and rainfall versus time (b) from May 1st 2004 to July 1st 2004. Details of soil suction versus time for rainfall periods (c, d, and e).**

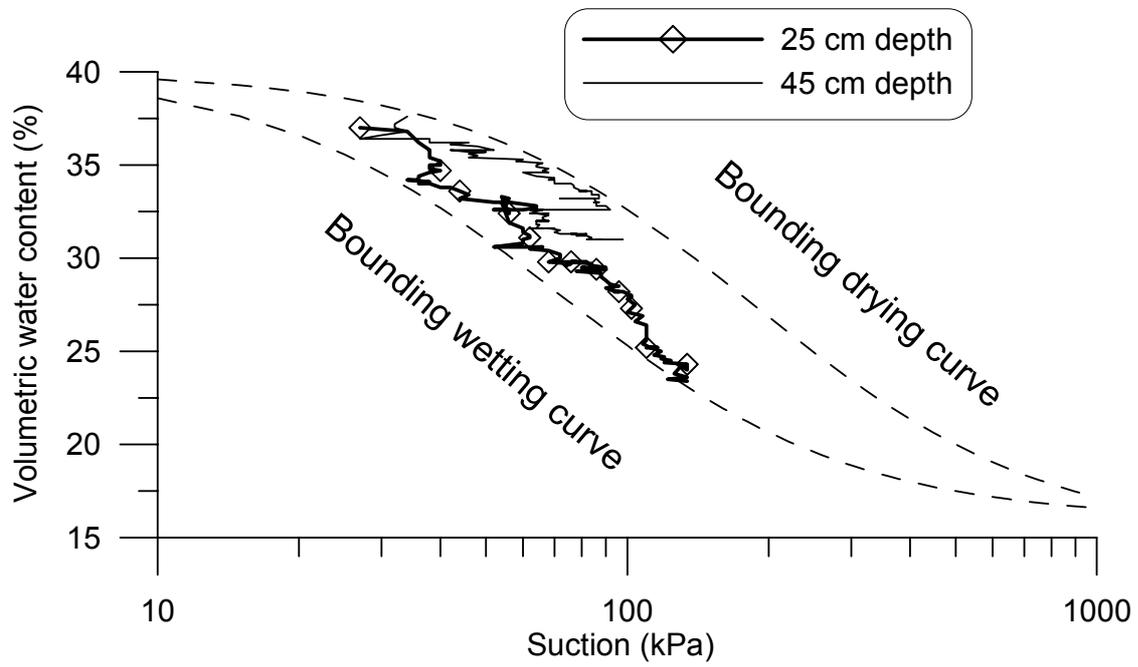

**Figure 4. Volumetric water content versus suction.**